\documentclass[prl,twocolumn,showpacs,preprintnumbers,amsmath,amssymb,a4paper]{revtex4-1}
\usepackage{graphicx}
\begin{document}

\title{Exploring multi-stability in semiconductor ring lasers: theory and experiment}

\author{L.~Gelens$^{1}$, S.~Beri$^{1,2}$, G.~Van~der~Sande$^1$, G.~Mezosi$^3$, M.~Sorel$^3$, J.~Danckaert$^{1,2}$ and G.~Verschaffelt$^1$}
\affiliation{
$^1$Department of Applied Physics and Photonics, Vrije Universiteit Brussel,
Pleinlaan 2, 1050 Brussels, Belgium;\\
$^2$Department of Physics, Vrije Universiteit Brussel,
Pleinlaan 2, 1050 Brussels, Belgium;\\
$^3$Department of Electronics \& Electrical Engineering, University of Glasgow, Rankine Building, Oakfield Avenue, Glasgow, G12 8LT, United Kingdom.
}

\date{\today}


\begin{abstract} 
We report the first experimental observation of multi-stable states in a single-longitudinal mode semiconductor ring laser. We show how the operation of the device can be steered to either monostable, bistable or multi-stable dynamical regimes in a controlled way. We observe that the dynamical regimes are organized in well reproducible sequences that match the bifurcation diagrams of a two-dimensional model. By analyzing the phase space in this model, we predict how the stochastic transitions between multi-stable states take place and confirm it experimentally. 
\end{abstract}
\maketitle

Multi-stability is a general feature of nonlinear systems which attracts attention in a broad set of subjects including hydrodynamics \cite{Ravelet_PRL_2004}, plasma physics \cite{Rempel06a}, biology \cite{Ozbudak_Nature_2004}, neural networks \cite{Foss_PRL_1996,Timme02a}, chemical reactions \cite{Chie_ChemPhysLett_2001} and optical systems \cite{Arecchi_PRL_1982,Xia07a,Agarwal82a}.
The phase space of a multi-stable system is in general very intricate due to the strongly interwoven basins of attractions of the coexisting stable structures, and is often further complicated by the presence of structures such as chaotic saddles \cite{Shrimali08a,Kraut02a}.
For this reason, the dynamics of a multi-stable system is characterized by a larger complexity than their bistable counterpart, leading to phenomena such as attractor hopping \cite{Kraut02a,Huerta_PRE_2008} or chaotic itineracy \cite{Timme02a}.
While being of broad interdisciplinary interest, multi-stability is especially interesting in the case of semiconductor lasers, 
due to their large number of applications and their wealth of dynamical regimes (see \cite{Wieczorek99a} and reference therein).
However, the fast time scales involved, the presence of spontaneous emission of photons blurring off the structures, the difficulty to control the internal parameters, the inaccessibility of some dynamical variables, all make the experimental reconstruction of the phase space of semiconductor lasers an extremely challenging task.
Therefore, despite the large number of theoretical work  \cite{Khovanov_PRL_2006,Khovanov00a,Kraut02a,Wieczorek99a,DeRossi98a,Agarwal82a}, the dynamical complexity of multi-stable semiconductor lasers remains experimentally unaddressed.\\
In this contribution, we experimentally address the phase space of Semiconductor Ring Lasers (SRLs), which are a novel class of semiconductor lasers characterized by circular geometry \cite{SorelOL2002}. We have focused our investigation on SRLs for several reasons. From the theoretical point of view, many dynamical regimes including multi-stable ones have been predicted \cite{Gelens_PRE_2009} but not observed yet.
From the technological point of view, an understanding of the phase space of SRLs would be highly desirable to
improve performances in applications such as all-optical memories \cite{HillNature2004} and allow for a better control of the device \cite{Khovanov00a,Lippi00a}.\\ 
We start our investigation with a comprehensive model \cite{VanderSandeJPhysB2008}, which allows us to predict bifurcation sequences and the different possible phase portraits of SRL-dynamics, including multi-stable ones.
Experimentally, we show how to control the parameters of a real SRL fully exploring its parameter space and the corresponding phase space portraits. 

\begin{figure}[t!]
\centering
\includegraphics[]{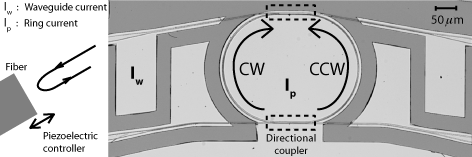}
\caption{\label{Fig:set_up}
Experimental SRL set-up. Four waveguide contacts are depicted of which only the one denoted by $\text{I}_{\text{w}}$ is biased. 
}
\end{figure}

Consider a SRL operating in a single longitudinal mode. Two directional modes with intensities $P_{1,2}$ and phases $\phi_{1,2}$ can propagate in the ring cavity. For sake of clarity, we assume that $P_1$ is the clockwise ($CW$) propagating mode and $P_2$ the counterclockwise one ($CCW$) [see Fig.\ \ref{Fig:set_up}].
The coupling between $CW$ and $CCW$ is described by a complex coupling parameter with amplitude $K$ and phase $\phi_k$ \cite{SorelOL2002}. On time scales slower than the relaxation oscillations, the total power $P_1+P_2$ is conserved and the dynamics of the SRL can be described by the following two-dimensional asymptotic model \cite{VanderSandeJPhysB2008}: 
\begin{eqnarray}
\dot{\theta} &= -2\sin\phi_k\sin\psi + 2 \cos\phi_k\cos\psi\sin\theta \nonumber \\
& + J \sin\theta\cos\theta , \label{Eq::Theta}\\
\cos\theta \dot{\psi} &= \alpha J \sin\theta\cos\theta + 2 \cos\phi_k \sin\psi \nonumber \\
&+ 2 \sin\phi_k \cos\psi \sin\theta. \label{Eq::Psi}
\end{eqnarray}
where $\theta = 2 \arctan \sqrt{P_2 / P_1} - \pi/2 \in [-\pi/2,\pi/2]$ 
represents the partitioning of power between modes, and $\psi = \phi_2-\phi_1 \in [0,2\pi]$ is the phase difference between the counter-propagating modes. 
$J$ is the rescaled bias current and $\alpha$ is the linewidth-enhancement factor. A full bifurcation analysis of Eqs.~(\ref{Eq::Theta})-(\ref{Eq::Psi}) has previously revealed a wide range of dynamical states \cite{Gelens_PRE_2009}. We focus here on two different values of the parameter $\phi_k$ representative for yet unexplored sequences of dynamical regimes including multi-stable ones.

\begin{figure}[t!]
\centering
\includegraphics[width=\columnwidth]{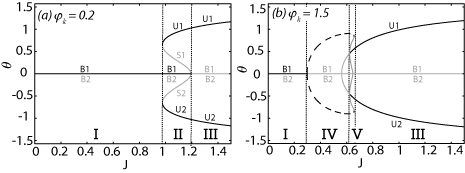}
\caption{\label{Fig:Phase_Space_Bif}
Bifurcation diagrams of Eqs.~(\ref{Eq::Theta})-(\ref{Eq::Psi}) for (a) $\phi_k = 0.2$ and (b) $\phi_k = 1.5$  
Stable solutions of Eqs.~(\ref{Eq::Theta})-(\ref{Eq::Psi}) are marked in black, unstable ones in gray. Dashed lines are used to indicate periodic solutions.
The Roman numbers {\bf I} - {\bf V} indicate different dynamical regimes.
}
\end{figure}

\begin{figure}[t]
\centering
\includegraphics[width=\columnwidth]{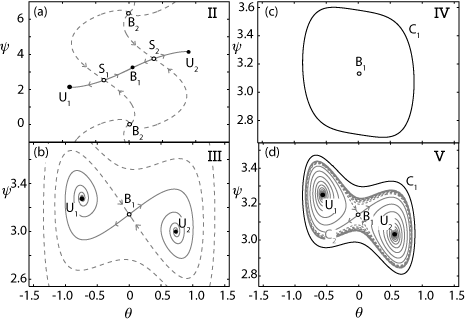}
\caption{\label{Fig:Different_PhaseSpacePics}
Phase space portraits of Eqs.~(\ref{Eq::Theta})-(\ref{Eq::Psi}) for different values of $J$ and $\phi_k$. The notation is as defined in the text. (a) $\phi_k = 0.2$, $J = 1.05$, (b) $\phi_k = 1.5$, $J = 0.75$, (c) $\phi_k = 1.5$, $J = 0.5$, (d) $\phi_k = 1.5$, $J = 0.66$.}
\end{figure}

\begin{figure}[t!]
\centering
\includegraphics[width=\columnwidth]{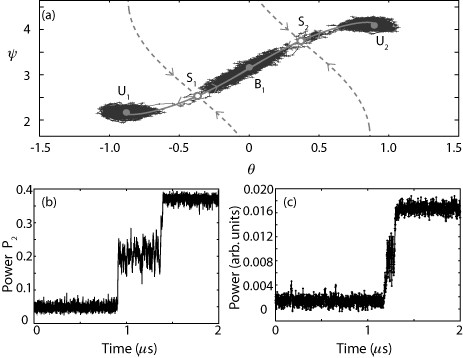}
\caption{\label{Fig:J_1_05_a_3_5_phi_0_2_Manifolds_OneSwitch} 
Simulations vs. experiments for a sequential transition $U_1 \to B_1 \to U_2$ for the phase space described in Fig.~\ref{Fig:Different_PhaseSpacePics}(b). 
(a) projection on the phase space; 
(b) numerical time series; (c) experimental data.
The model parameters for (a)-(b) are $J=1.05$, $\phi_k=1.02$, $\alpha=3.5$.
The experimental conditions for (c) were $I_p = 45.31$mA and $I_w = 9.81$mA.
}
\end{figure}

Consider $\phi_k = 0.2$. The bifurcation diagram for such value of $\phi_k$ is shown in Fig.\ \ref{Fig:Phase_Space_Bif}(a) 
while representative examples of its phase space are shown in Fig.\ref{Fig:Different_PhaseSpacePics}.
For small values of the bias current, the system operates in bidirectional regime and the phase space of the system (not shown) consists of a stable state $B_1$ coexisting with an unstable one $B_2$ (Region $\bf I$).
An increase of the bias current $J$ leads to the appearance of two more stable states corresponding to unidirectional $CW$ and $CCW$ rotating solutions $U_{1,2}$. The SRL therefore operates in a tristable regime (Region $\bf II$). The corresponding phase space is shown in Fig.~\ref{Fig:Different_PhaseSpacePics}(a). Three stationary states $U_1$, $U_2$ and $B_1$ coexist. An unstable state $B_2$ corresponding to bidirectional in-phase lasing and two saddles $S_{1,2}$ are also present in the system. The basins of attraction of the three states are separated by the stable manifolds of $S_1$ and $S_2$ in such a way that the basin of attraction of $B_1$ lies in between the basins of attraction of $U_1$ and $U_2$. 
When spontaneous emission noise is introduced in the system, spontaneous attractor hopping may appear. In the limit of vanishing noise intensity, the topology of the phase space predicts only transitions between a unidirectional mode and the bidirectional mode. Direct transitions between $U_1$ and $U_2$ are possible only for larger values of the noise intensity. An example of a simulated hopping trajectory is shown in Fig.\ \ref{Fig:J_1_05_a_3_5_phi_0_2_Manifolds_OneSwitch}(a)-(b), and will be compared with experimental switches later. Increasing the value of the bias current $J$, the saddles $S_{1,2}$ migrate towards $B_1$ making the basin of attraction of $B_1$ shrink. Therefore the residence time in the bidirectional lasing mode is expected to decrease and the laser to operate most of its time in either one of the unidirectional modes. When the bias current is increased above a critical value, the saddles collide with $B_1$ and the bidirectional out-of-phase mode become unstable. The SRL is then operating in a bistable regime with two stable unidirectional modes $U_{1,2}$ (Region $\bf III$). The corresponding phase space is shown in Fig.~\ref{Fig:Different_PhaseSpacePics}(b). Here $U_1$ and $U_2$ are the unidirectional modes whereas $B_1$ is a saddle point. The stable manifold of $B_1$ separates the basins of attraction of $U_1$ and $U_2$.
In this regimes, noise-induced hopping is expected between the two unidirectional modes, whereas we do not expect to observe any residence in the bidirectional regime. 
A further increase of the current leads to longer residence times in $U_{1,2}$ but no further bifurcations are expected.

When $\phi_k=1.5$ the bifurcation curve as shown in  Fig.~\ref{Fig:Phase_Space_Bif}(b) is qualitatively different from the previous case. For small values of the bias current, the system operates in the bidirectional regime, similar to Region $\bf I$ for $\phi_k=0.2$. Increasing the current above a critical value, the 
bidirectional operation loses its stability and the SRL exhibits periodic oscillations known as alternate oscillations \cite{SorelOL2002} between $CW$ and $CCW$ modes (Region $\bf IV$).
In the phase space [Fig.~\ref{Fig:Different_PhaseSpacePics}(c)] the alternate oscillations are a stable limit cycle $C_1$ which surrounds the unstable bidirectional state $B_1$. When the bias current $J$ is increased, two unidirectional solution $U_{1,2}$ appear, and tristability between the latter and $C_1$ is possible (Region $\bf V$). This scenario corresponds to Fig.~\ref{Fig:Different_PhaseSpacePics}(d). The basins of attractions  of $U_{1,2}$ are separated by the stable manifold of the saddle $B_1$. A second unstable cycle $C_2$ separates the basin of attraction of $C_1$ from the basins of attraction of $U_{1,2}$. When noise is present in the system hopping between $U_1$ and $U_2$ as well as hopping between $U_{1,2}$ and the alternate oscillations are possible, allowing the system to burst into periodic oscillations.
A further increase of the current leads to the disappearance of $C_1$ and the SRL operates in a bistable regime between the two unidirectional modes $U_{1,2}$ (Region $\bf III$).
The phase space corresponding to this regime is shown in Fig.~\ref{Fig:Different_PhaseSpacePics}(b), similar to the case of $\phi_k=0.2$. The stochastic terms induce hopping between $U_1$ and $U_2$, but no periodic oscillations can appear.
 
The experiments have been performed on an InP-based multiquantum-well SRL with a racetrack geometry and a free-spectral-range of $53.6$ GHz. The experimental set-up is shown in Fig.\ \ref{Fig:set_up}. The device operates in a single-transverse, single-longitudinal mode regime at wavelength $\lambda = 1.56 \mu$m. 
The chip containing the SRL is mounted on a copper mount and thermally controlled by a Peltier element which is stabilized at a temperature of $28.57 ^\circ$C with an accuracy of $0.01^\circ$C.
A bus waveguide made of the same active material as the ring has been integrated on the chip in order to couple power out from the ring. To this waveguide, an independent electrical contact has been applied. Sending current through the waveguide reduces the absorption. The power emitted from the chip is collected with a multimode fiber and detected with a $2.4$GHz photodiode connected to an oscilloscope.
The strength $K$ and phase $\phi_k$ of the coupling between $CW$ and $CCW$ modes are not controllable during the fabrication process and they are {\it a priori} unknown.
However, by using the cleaved facet of the fiber as a mirror, we are able to reflect power from one mode (for instance $CCW$) back into the waveguide and finally to the counterpropagating mode in the ring. 
The amount of power that is coupled to the $CW$ mode can then be controlled by tuning the current $I_w$ on the waveguide, whereas its phase can be tuned by positioning the fiber facet with a piezoelectric controller. With this technique, we have achieved full control of the coupling parameter $\phi_k$ as well as the coupling strength $K$.

\begin{figure}[t!]
\centering
\includegraphics[width=\columnwidth]{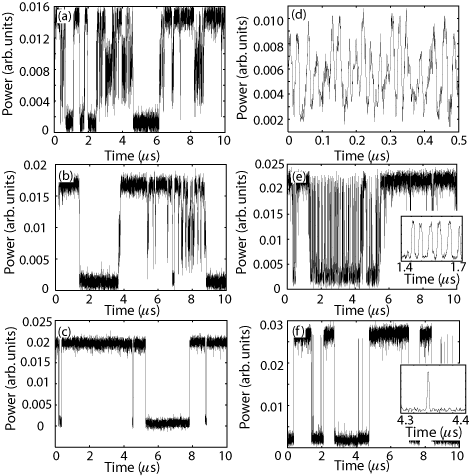}
\caption{\label{Fig:timetraces} 
Experimentally measured time series for different bias currents $I_p$ and $I_w$ corresponding to the dynamical regimes of Fig.~\ref{Fig:Different_PhaseSpacePics}. 
(a) $I_w$ = $9.81$mA, $I_p$ = $44.35$mA;
(b) $I_w$ = $9.81$mA, $I_p$ = $45.21$mA;
(c) $I_w$ = $9.81$mA, $I_p$ = $46.39$mA;
(d) $I_w$ = $12.0$mA, $I_p$ = $40.23$mA;
(e) $I_w$ = $12.0$mA, $I_p$ = $44.34$mA;
(f) $I_w$ = $12.0$mA, $I_p$ = $45.94$mA.
The insets show a zoom on relevant segments of the time series.
}
\end{figure}

We have fixed the waveguide current $I_w = 9.81$mA and tuned the voltage on the piezoelectric controller until a symmetric operation of the SRL was achieved. 
According to the discussion above, this corresponds to fixing the $\phi_k$ parameter of the SRL. We now tune the bias current $I_p$ on the ring, in order to reproduce the dynamical regimes predicted by the theory.
We choose here to measure with the oscilloscope the emission in the $CCW$ mode; due to the anti-correlated dynamics of the counterpropagating modes \cite{SorelOL2002}, any change in the power in the $CCW$ mode corresponds to an opposite change in the emission in the $CW$ mode.
A high (low) amplitude signal on the scope thus corresponds to operation in the $CCW$ ($CW$) mode, whereas bidirectional operation appears as an intermediate amplitude signal.
Examples of timetraces of the $CCW$ mode are shown in Fig.~\ref{Fig:timetraces} (a)-(f). 
The ring reaches its lasing threshold at $34$mA. For current values close to threshold, bidirectional lasing is observed (not shown).
When increasing the current above a critical value $I_p \sim 44$mA, hopping between the bidirectional regime and the two unidirectional modes appear. 
Segments of time traces for  $I_p = 44.35$mA and $I_p = 45.21$mA are shown in Fig.~\ref{Fig:timetraces}(a)-(b). The average residence time in the $CW$ and $CCW$ state increases with the pump current, while the residence time in the bidirectional mode decreases.
Tristability between bidirectional and unidirectional modes is observed. In agreement with the phase space picture in Fig.~\ref{Fig:Phase_Space_Bif}(a), hopping events preferentially occur between the bidirectional and one of the unidirectional modes. A detail of a sequential transition from the $CW$ mode to bidirectional operation, to the $CCW$ mode for $I_p = 45.21$mA is shown in Fig.~\ref{Fig:J_1_05_a_3_5_phi_0_2_Manifolds_OneSwitch}(c). The agreement with the numerical simulations of Eqs.~(\ref{Eq::Theta})-(\ref{Eq::Psi}) as shown in Fig.~\ref{Fig:J_1_05_a_3_5_phi_0_2_Manifolds_OneSwitch}(b) is clear. The observation of some direct transitions between unidirectional modes suggests that the noise-induced diffusion length is not negligible when compared to the size of the basin of attraction of the bidirectional mode. When the pump current is increased to $I_p = 46.39$mA as shown in Fig.~\ref{Fig:timetraces}(c), no bidirectional operation is observed, and direct transitions between $CW$ and $CCW$ modes are possible as predicted by the phase space portrait in Fig.~\ref{Fig:Different_PhaseSpacePics}(b).
Further increase in the bias current correspond to an increase in the average residence time in the two unidirectional modes.
As such the whole bifurcation sequence as in Fig.~\ref{Fig:Phase_Space_Bif}(a) has been experimentally reconstructed.

We then fix the waveguide current to $I_w = 12.0$mA and we adjust the voltage on the piezoelectric until the symmetry in the system is restored.
Once again we tune the pump current $I_p$ on the ring and we investigate the different dynamical regimes. Typical time traces for the $CCW$ mode are shown in Fig.~\ref{Fig:timetraces}(d)-(f).
Close to threshold the laser operates in bidirectional regime, for slightly higher values of the bias current, alternate oscillations appear as reported in Fig.~\ref{Fig:timetraces}(d) for $I_p = 40.23$mA. This operating regime reveals the phase space portrait shown in Fig.~\ref{Fig:Different_PhaseSpacePics}(c). When increasing the bias current, the amplitude of the alternate oscillations increases until the SRL becomes multi-stable and 
the alternate oscillations coexist with the two unidirectional modes as shown in Fig.~\ref{Fig:timetraces}(e) for a bias current $I_p = 44.35$mA.
The inset in Fig.~\ref{Fig:timetraces}(e) is a zoom on a burst that reveals the periodical oscillations. This operating regimes corresponds to the phase space portrait shown in Fig.\ref{Fig:Different_PhaseSpacePics}(d).
In this regime both transitions between the unidirectional modes and between the unidirectional modes and the limit cycle are observed, as allowed by the topology of the phase space in  Fig.\ref{Fig:Different_PhaseSpacePics}(d).
When the current is further increased, the bursts of oscillations disappear and bistability between $CW$ and $CCW$ modes is achieved as shown in Fig.~\ref{Fig:timetraces}(f) for $I_p = 45.94$mA.
Such regimes corresponds to the phase space pictured in Fig.\ref{Fig:Different_PhaseSpacePics}(b).
Short excursions from $CW$ to $CCW$ operation and vice versa are observed in the time traces when the laser operates in this regime [see inset in Fig.~\ref{Fig:timetraces}(f)]. They have been previously observed \cite{Beri_PRL_2008} and explained as noise induced diffusion between the folds of the stable manifolds of the saddle point $B_1$ in Fig.~\ref{Fig:Different_PhaseSpacePics}(b).
The presence of such excursions in the time trace of Fig.~\ref{Fig:timetraces}(f) represents a further confirmation of the phase space structure described in Fig.~\ref{Fig:Different_PhaseSpacePics}(b).

In conclusion, we have performed a comprehensive experimental investigation of the phase space of a SRL. The dynamical regimes that we revealed experimentally match the phase space topologies of Eqs.~(\ref{Eq::Theta})-(\ref{Eq::Psi}). A control scheme based on the reflection of power from the cleaved facet of an optical fiber and the active bias of the bus waveguide has been devised in order to control the (otherwise unaccessible) coupling parameter $\phi_k$. In this way, we could explore the dynamics of the SRL over the whole parameter space $J-\phi_k$, including previously undisclosed regimes. Whereas alternate oscillations, bidirectional and unidirectional operation have previously been reported in SRLs \cite{SorelOL2002}, Fig.~\ref{Fig:timetraces}(e) represents the first experimental observation of the coexistence between alternate oscillations and unidirectional operation, and more generally shows coexistence of a limit cycle and two stable nodes in a semiconductor laser system. Moreover, coexistence of three stable nodes has been demonstrated in Fig.~\ref{Fig:timetraces}(a)-(b).Our experiments have been performed on a SRL, and we expect our findings to be extendable to other systems that display $Z_2$ symmetry, such as for instance disk-lasers.

This work has been partially funded by the European Union under project IST-2005-34743 (IOLOS). This work was supported by the Belgian Science Policy Office under grant No.\ IAP-VI10. We acknowledge the Research
Foundation-Flanders (FWO) for individual support and project funding.


\end{document}